\documentclass[3p,times,twocolumn]{elsarticle}

\usepackage{ecrc}

\volume{00}

\firstpage{1}

\journalname{Nuclear Physics B Proceedings Supplement}

\runauth{}

\jid{nuphbp}

\jnltitlelogo{Nuclear Physics B Proceedings Supplement}

\usepackage{amssymb}
\usepackage{overpic}
\usepackage{booktabs}

\usepackage{lineno}

\usepackage[figuresright]{rotating}

\begin{document}

\begin{frontmatter}



\dochead{}

\title{The artificial retina for track reconstruction at the LHC crossing rate}


\author[milano]{A.~Abba}
\author[infn]{F.~Bedeschi}
\author[milano]{M.~Citterio}
\author[milano]{F.~Caponio}
\author[milano]{A.~Cusimano}
\author[milano]{A.~Geraci}   
\author[sns,infn]{P.~Marino\corref{corrauth}}
\ead{pietro.marino@pi.infn.it}
\cortext[corrauth]{Corresponding author.}
\author[sns,infn]{M.~J.~Morello}
\author[milano]{N.~Neri}
\author[unipi,infn]{G.~Punzi}
\author[unipi]{A.~Piucci}
\author[infn,fermilab]{L.~Ristori}
\author[infn]{F.~Spinella}
\author[sns,infn]{S.~Stracka}
\author[cern]{D.~Tonelli}

\address[unipi]{University of Pisa,
 Lungarno Pacinotti 43, 56126, Pisa, Italy}
 \address[sns]{Scuola Normale Superiore,
Piazza dei Cavalieri 7, 56127, Pisa, Italy}
\address[infn]{INFN-Pisa,
 L.go Bruno Pontecorvo 3, 56127, Pisa, Italy}
 \address[milano]{Politecnico and INFN-Milano,
 Via Celoria 16, 20133, Milano, Italy}
 \address[fermilab]{Fermilab,
Wilson and Kirk Rd, Batavia, IL 60510, USA}
\address[cern]{CERN
  385 Route de Meyrin, Geneva, Switzerland}

\begin{abstract}
We present the results of an R\&D study for a specialized processor 
capable of precisely reconstructing events with hundreds of charged-particle 
tracks in pixel and silicon strip detectors at $40\,\rm MHz$,
 thus suitable for processing LHC events at the full crossing frequency. 
 For this purpose we design and test a massively parallel pattern-recognition algorithm, 
 inspired to the current understanding of the mechanisms adopted 
 by the primary visual cortex of mammals in the early stages of visual-information processing.
The detailed geometry and charged-particle's activity of a large tracking detector 
are simulated and used to assess the performance of the artificial retina algorithm. 
We find that high-quality tracking in large detectors is possible with sub-microsecond latencies 
when the algorithm is implemented in modern, high-speed, high-bandwidth FPGA
devices.
\end{abstract}


\end{frontmatter}


\section{Introduction}
\label{sec:intro}

Higher LHC energy and luminosity increase the challenge 
of data acquisition and event reconstruction in the LHC experiments.
The large number of interactions for bunch crossing (pile-up) greatly reduces
the discriminating power of usual signatures, 
such as the high transverse momentum of leptons or the high transverse missing energy.
Therefore real-time track reconstruction could prove crucial to quickly select
potentially interesting events for higher level of processing.
Performing such a task
at the LHC crossing rate is a major challenge
because of the large combinatorial and the size of the associated information flow
and requires unprecedented massively parallel pattern-recognition algorithms.
For this purpose we design and test a  
neurobiology-inspired pattern-recognition algorithm 
well suited for such a scope: the \emph{artificial retina} algorithm.

\section{An \emph{artificial retina} algorithm }\label{sec:retina}

The original idea of an artificial retina tracking algorithm was inspired
by the mechanism of visual receptive fields in the mammals eye~\cite{RistoriNIM}.
Experimental studies have shown neurons
tuned to recognize a specific shape on specific region of the retina
(``receptive field'')
%
The strength of the response of each neuron to a stimulus is proportional to how 
close 
the shape of the stimulus is to the shape for which the neuron is tuned to. 
All neurons react to a stimulus, each with different strength, 
and the brain obtains a precise information of the received stimulus 
performing some sort of interpolation between the responses of neurons.

The retina concepts
can be geared toward track reconstruction.
Assuming a generic tracking detector, 
the 3D charged particle trajectory is described by five parameter.
The space of track parameters are discretized into \emph{cells}, which mimic the
receptive fields of the retina.
The center of each cell identifies a track in the detector space, that intersects
detector layers in spatial points that we call \emph{receptors}. 
For each incoming hit, the algorithm computes
the excitation intensity, \emph{i.~e.} the response of the receptive field,
of each cell as follows:
\begin{equation}
\label{eq:responsecell}
R = \sum_{k,\,r} \exp\Big(-\frac{s^2_{kr}}{2\sigma^2}\Big),
\end{equation}
where $s_{kr}$ is the distance, on the layer $k$, between the hit and the receptor $r$.
$\sigma$ is a parameter of the retina algorithm, that can be adjusted 
to optimize the sharpness of the response of the receptors.

After all hits are processed, tracks are identified as local maxima 
over a threshold in the space of track parameters.
Averaging over nearby cells of the identified maximum provides
track parameters 
with a significant better resolution
than the available cell granularity. 

\section{Retina algorithm in a real HEP experiment}

To evaluate the performances and the robustness of the algorithm in a real
HEP detector,
we focus on the upgraded LHCb detector.
The upgraded LHCb detector~\cite{LHCb_TDR}, a single-arm spectrometer covering the pseudo-rapidity 
range $2<\eta<5$, is a major upgrade
of the current LHCb experiment, and it will run  
at the instantaneous luminosity of $3\times 10^{33}$~cm$^{-2}$s$^{-1}$, with 
a beam energy of $7\,\mathrm{TeV}$.
%
%
%
%
All the sub-detectors will be read out at $40\,\rm MHz$, 
allowing a complete event reconstruction at the LHC crossing rate.
To benchmark the retina algorithm, we decided
to perform the first stage of the upgraded LHCb detector tracking reconstruction~\cite{VELOUT},
using the information of 
only two sub-detectors, placed upstream of the magnet:
the vertex locator (VELO), a silicon-pixel detector~\cite{VELOTDR} and
the upstream tracker (UT)~\cite{UTTDR}, a silicon microstrip detector.
We used the last eight forward pixel layers of the 
VELO and the two axial layers of the UT.
We arbitrarily chose to parametrize tracks with the following parameters: $(u,v, d,z_0, k)$.
$(u,v)$ are the 
the spatial coordinate of the intersection point of the track with a ``virtual plane''
perpendicular to the $z$-axis, placed to a distance $z_{vp}$ from the origin of the 
coordinate system. 
%
%
$d$ is the signed transverse impact parameter, $z_0$ is the
$z$-coordinate of the point of the closest approach to the $z$-axis.
$k$ is the signed curvature in the bending plane $(\vec{B} = B\hat{y})$.

The detector geometry and magnetic field (negligible in the VELO and about $0.05\,\rm T$ in the UT),
allow us to use only the $(u,v)$ parameters to perform the pattern recognition, 
since the 5D tracks' parameters space can be factorized into $(u,v) \otimes (d, z_0, k)$. 
Thus $(u,v)$ are the ``main'' parameters where pattern recognition is performed, 
whereas $(d,z_0,k)$ are treated as ``perturbation'' of the main parameters 
$(u,v)$~\cite{NotaPubLHCb, Jinst_Punzi}.

To evaluate the performances of the algorithm, we develope
a detailed \texttt{C++} simulation of the retina algorithm~\cite{WIT_Marino} 
able to process simulated events,
interfaced with the default LHCb simulation.
We discretize the main $(u,v)$-subspace into $22\,500$ cells, 
a granularity $\mathcal{O}(100)$
larger than the maximum expected number of tracks in a typical upgraded LHCb event.
\begin{figure}[tbp]
\centering
\includegraphics[width=.52\columnwidth]{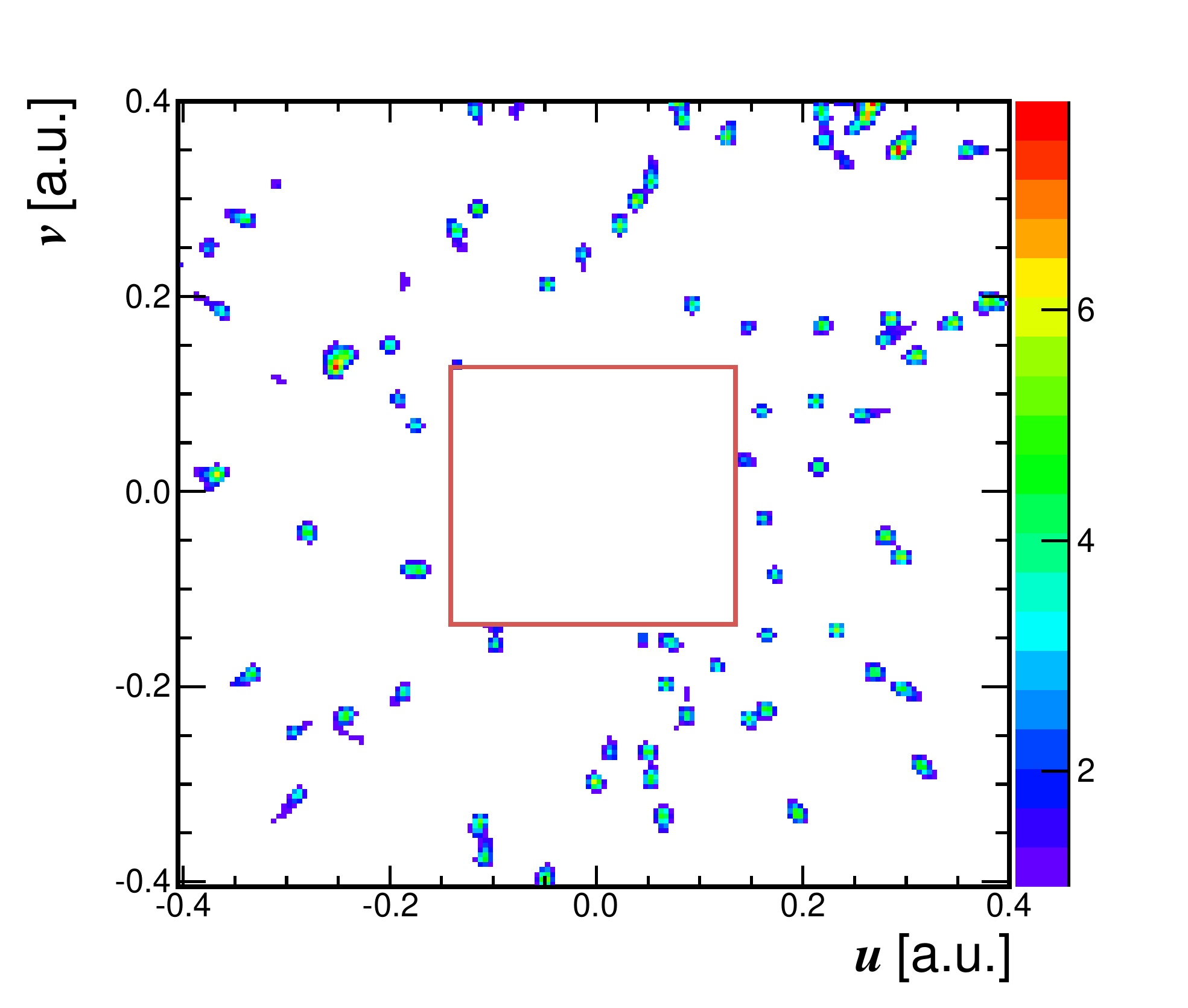} 
\includegraphics[width=.46\columnwidth]{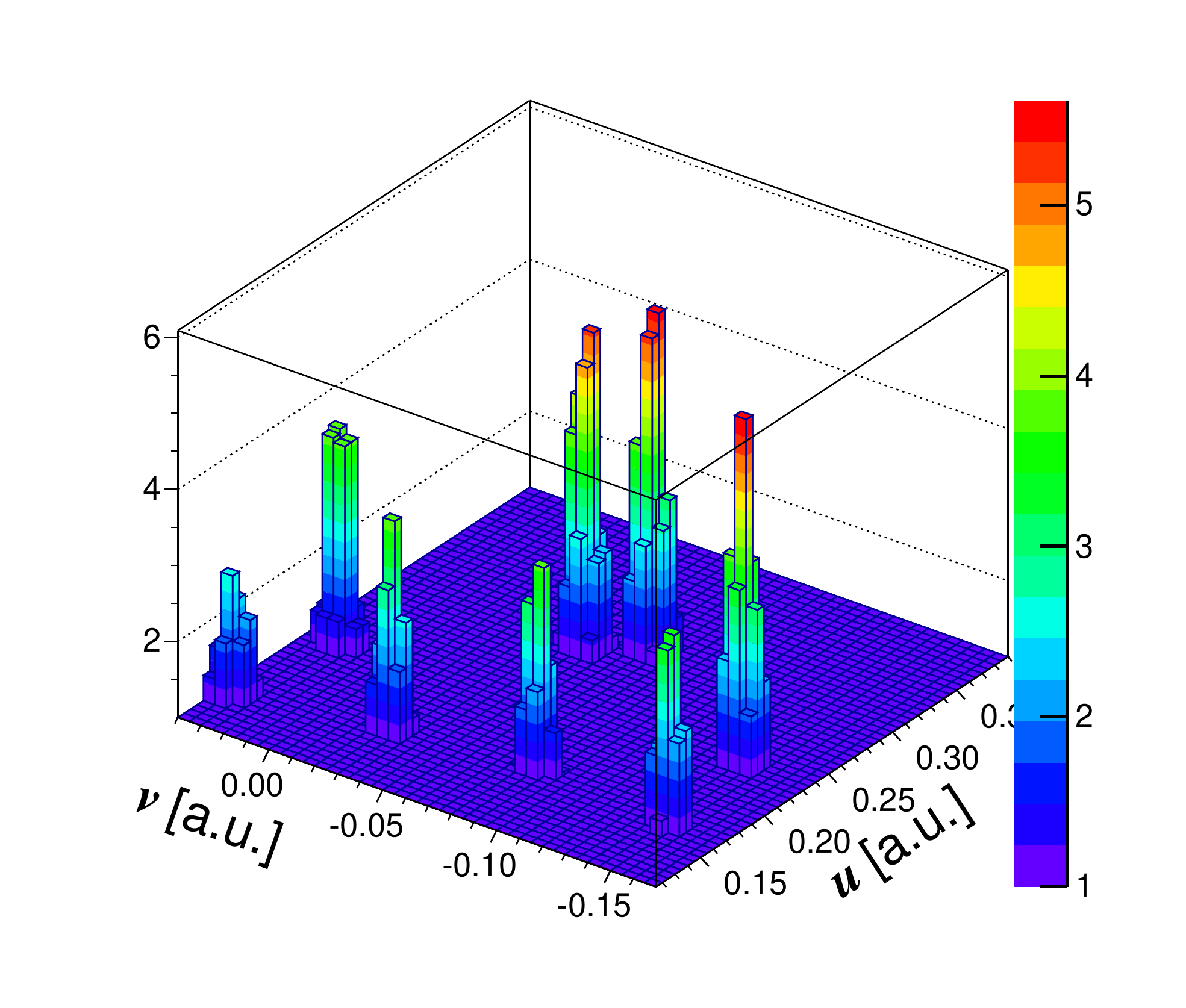} 
\caption{Left: response of the retina algorithm
(only the $(u,v)$-plane, where the pattern recognition is made) to
a generic collision from the default LHCb simulation, 
with instantaneous luminosity of $L=2\times 10^{33} \, \rm cm^{-2} s^{-1}$. 
The hole at the center of the figure is due to the physical hole in the VELO layers.
Right: a zoom of the retina response.}
\label{fig:ev_MC}
\end{figure}
Generic collisions samples from the default LHCb simulation are used to assess the performances
of the retina algorithm.
The generic collisions are generated 
with beam energy of $7\,\mathrm{TeV}$, 
and luminosities up to 
$L=3\times 10^{33} \, \rm cm^{-2} s^{-1}$.
A typical response of the retina algorithm 
is shown in fig.~\ref{fig:ev_MC}, where 
several clusters are clearly identifiable, 
and most of them reconstructed as tracks.

\begin{figure}[tbp]
\centering
\begin{overpic}[width=.492\columnwidth]{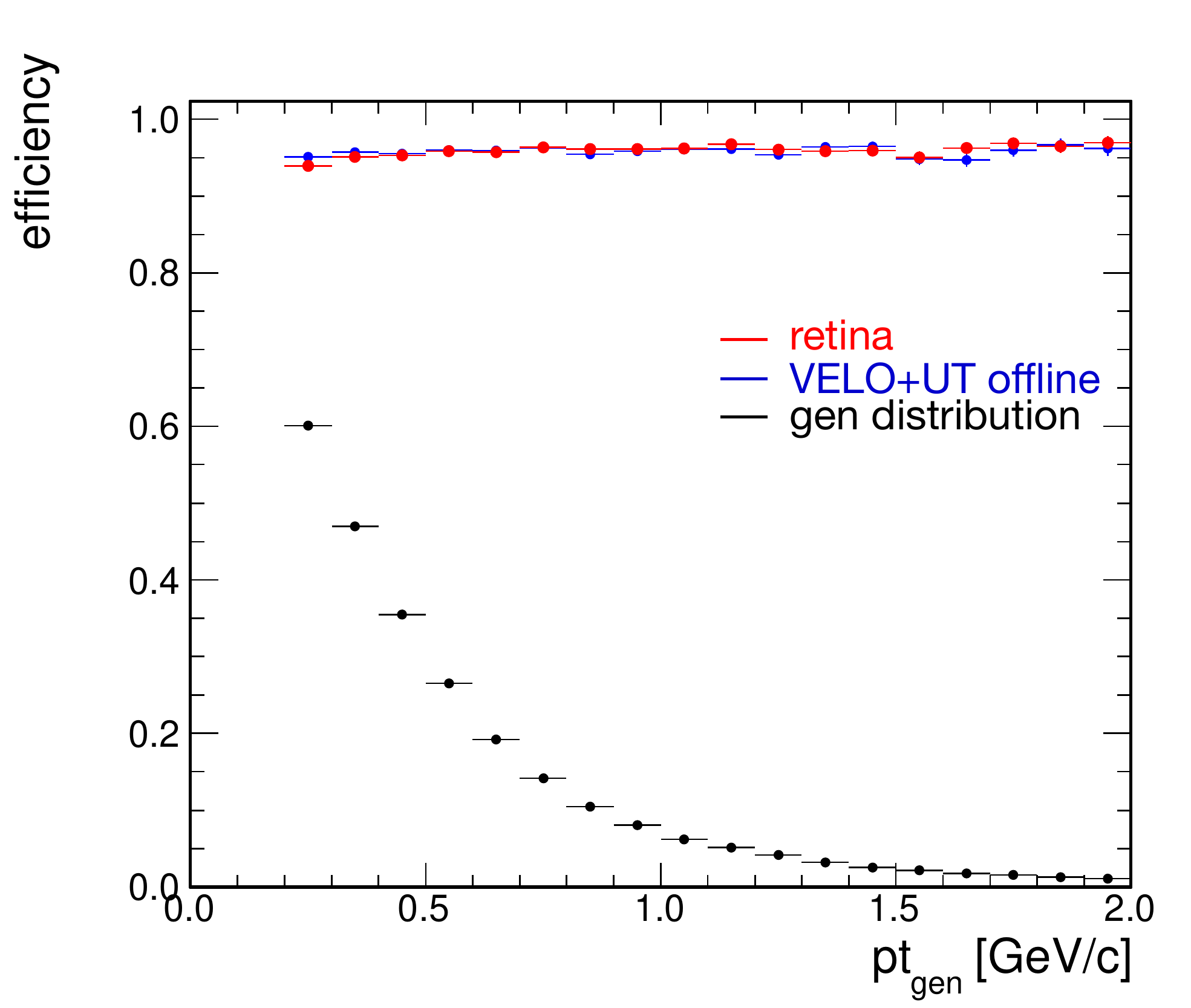}
	\put(75, 35){$(a)$}
\end{overpic} 
\begin{overpic}[width=.492\columnwidth]{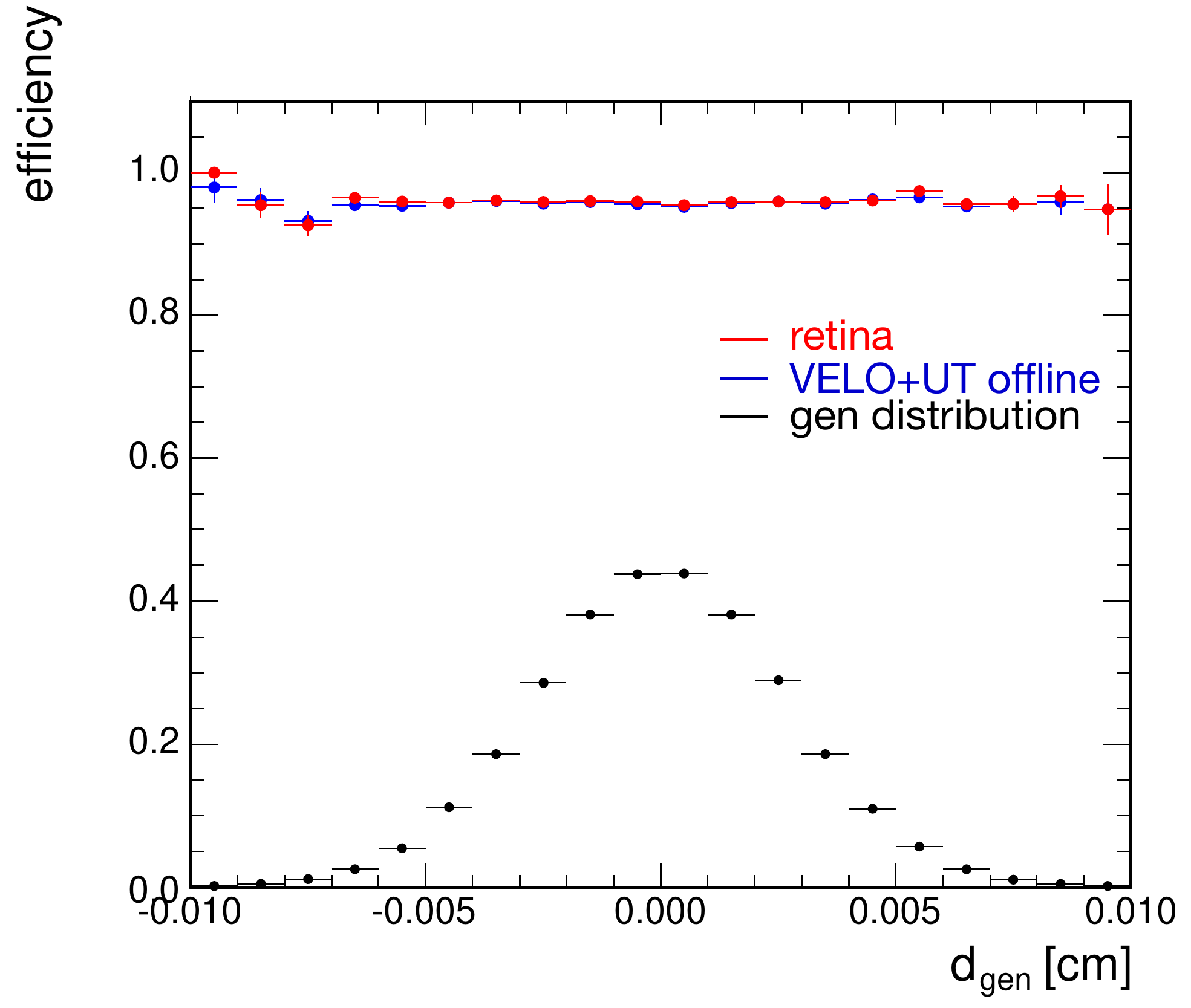}
	\put(75, 35){$(b)$}
\end{overpic} 
\caption{Tracking reconstruction efficiency of the retina algorithm (in red) and 
of the offline VELO+UT algorithm (in blue), as function
of:  $(a)$ $p_T$, $(b)$ $d$. 
The distribution of the considered 
parameter is, also, reported in black.
Luminosity of $L=3\times 10^{33} \, \rm cm^{-2} s^{-1}$.}
\label{fig:efficiency_retina}
\end{figure}

All hits from simulated
events from the default LHCb simulation are sent and processed by the retina.
In order to evaluate tracking performances we considered 
only tracks in a region of the $(u,v)$-plane
where they have full acceptance on the chosen layer configuration.
In addition, cuts close to the ones applied to calculate the offline efficiency~\cite{VELOUT}
are applied.
For instance, we required at least three hits on VELO layers
and two hits on UT layers, and also a momentum $p>3\,\mathrm{GeV}/c$ 
and a transverse momentum  $p_T>200\,\mathrm{MeV}/c$.
Tracks satisfying all these requirements are defined as
\emph{reconstructable},  and the 
tracking efficiency is defined as the number of reconstructed tracks 
over the number of reconstructable tracks. 
The efficiency of the retina is reported in figure~\ref{fig:efficiency_retina} 
as function of $p_T, \,d$ parameters. 
We also report the efficiency  of the offline LHCb
track reconstruction algorithm,
performing the same task as the retina~\cite{NotaPubLHCb}.
The retina algorithm shows very
high efficiencies in reconstructing tracks,  about 95\% for generic tracks, which is comparable to the
offline tracking algorithm.  The fake track rate is
8\% at $L=2\times 10^{33} \, \rm cm^{-2} s^{-1}$ and 12\% at $L=3\times 10^{33} \, \rm cm^{-2} s^{-1}$,
slightly higher than the fake rate of the offline algorithm. 
We also estimate the efficiency of the retina algorithm in
recostruncting signal tracks from some  benchmark decay modes, 
such as $B_s^0 \to \phi\phi$, $D^{*\pm} \to D^0\pi^{\pm}$ and $B^0 \rightarrow K^{*} \mu \mu$
for $L = 2\times 10^{33}\,\rm cm^{-2}s^{-1}$. The efficiency for these channels is about $97$--$98\%$.
Resolutions on tracking parameters determined by the retina are 
comparable with those of the offline reconstruction.

\section{Hardware implementation}
\begin{figure}[tb]
\centering
\includegraphics[width=.9\columnwidth]{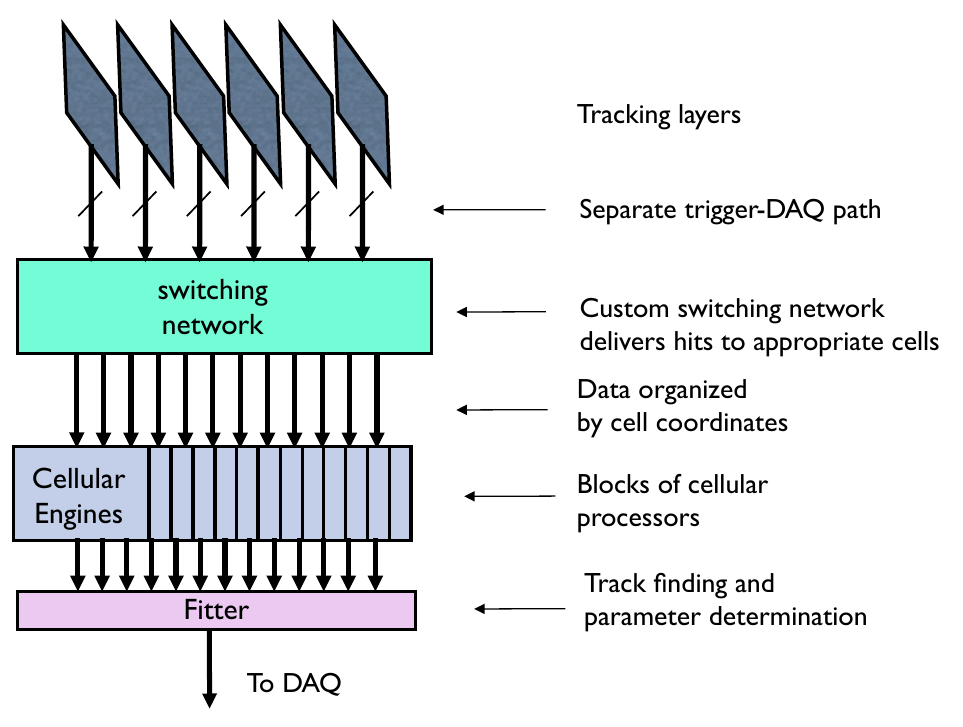}
\caption{Illustration of the device’s architetture.\label{fig:arch}}
\end{figure}

To fully exploit the high-grade of parallelism of the algorithm, 
we developed the retina algorithm into FPGA chips~\cite{WIT_Tonelli}.
The logic is implemented in VHDL language;
detailed logic-gate placement and simulation
on the high-bandwidth Altera Stratix V device model \texttt{5SGXEA7N2F45C2ES}
is achieved using Altera's proprietary software.
Figure~\ref{fig:arch} shows an overview of the device’s architetture.
To achieve an efficient distribution of the hit information 
coming from the detector layers to the cells of the space of track parameters, 
we design an intelligent information
delivery system that routes each hit in parallel to all and only those cells
for which such hit is likely to contribute a significant weight.
The switching network completes its processing in 30 clock cycles.
Each cell in the track’s parameter space is defined as a logic module, the
\emph{engine}. The engine is implemented as a clocked pipeline, that calculate the excitations.
The engine process takes 17 clock cycles.
At the end, the logic that identifies the center-of-mass in the space of track parameters
take 11 cycle of clock cycles along with another 10 cycles for fanout.
With a clock frequencies of $350\,\rm MHz$, the latency for reconstructing online tracks is less than 
$0.5\,\rm \mu s$.
Each Stratix V can host up to $900$ engines leaving approximately 25\% 
of logic available for other uses, including a 15\% of switching and the logic for center-of-mass 
calculation~\cite{NotaPubLHCb}.

\section{Conclusions}
We showed that high-quality tracking in large LHC detectors is possible at a $40\,\rm MHz$
event rate with sub-$\rm \mu s$ latencies, 
when appropriate parallel algorithms are used in conjunction with current high-end FPGA device.
This opens the interesting possibility of designing high-rate experiments where track reconstruction 
happen transparently as part of the detector readout.





\nocite{*}
\bibliographystyle{elsarticle-num}
\bibliography{ICHEP14}







\end{document}